\documentclass[a4paper,fleqn,usenatbib]{mnras}
\usepackage{amssymb,amsmath,graphicx,psfrag,mathtools}
\usepackage[T1]{fontenc} 
\usepackage{ae,aecompl} 
\usepackage{url}
\usepackage{revsymb}
\hypersetup{draft} 
\title[Outliers]{When Outliers Are Different}
\author[J. I. Katz]{
J. I. Katz,$^{1}$\thanks{E-mail katz@wuphys.wustl.edu} 
\\
$^{1}$Department of Physics and McDonnell Center for the Space Sciences,
Washington University, St. Louis, Mo. 63130 USA 
}
\date{Accepted XXX.  Received YYY; in original form ZZZ} 
\pubyear{2021} 
\date{\today}
\begin{document} 
\label{firstpage} 
\pagerange{\pageref{firstpage}--\pageref{lastpage}} 
\maketitle 
\begin{abstract}
	When does the presence of an outlier in some measured property
	indicate that the outlying object differs qualitatively, rather than
	quantitatively, from other members of its apparent class?  Historical
	astronomical examples include the many types of supernov\ae\ and
	short {\it vs.\/} long Gamma Ray Bursts.  A qualitative difference
	implies that some parameter has a characteristic scale, and hence
	its distribution cannot be a power law (that can have no such
	scale).  If the distribution is a power law the objects differ only
	quantitatively.  The applicability of a power law to an empirical
	distribution may be tested by comparing the most extreme member to
	its next-most extreme.  The probability distribution of their ratio
	is calculated, and compared to data for stars, radio and X-ray
	sources, and the fluxes, fluences and rotation measures of Fast
	Radio Bursts (FRB).  {It is found with high statistical significance
	that the giant outburst of soft gamma repeater SGR 1806-20 differed
	qualitatively from its lesser outbursts and FRB 200428 differed
	qualitatively from other FRB (by location in the Galaxy), but that
	in some supernova remnant models of rotation measure FRB 121102 is
	not, statistically significantly, an outlier.}
\end{abstract}
\section{Introduction}
A question often asked in astronomy is whether a group of objects, similar
in some ways, may be subdivided into qualitatively distinct subclasses.  For
example, nov\ae, originally considered a single class of events, were first
subdivided into Galactic nov\ae\ and supernov\ae\ (not observed in our Galaxy
since 1604, but recognized in the Andromeda Galaxy in 1885).  Subsequently,
supernov\ae\ have been divided into an ever-multiplying botanic garden of
types and subtypes.  Gamma-Ray Bursts (GRB) were divided into (Galactic and
Magellanic Cloud) Soft Gamma Repeaters and (extragalactic) GRB, and then GRB
into short and long GRB \citep{K93} that were found to differ not only in
duration but in physical origin.  Qualitative divisions have been essential
to understanding these phenomena, because no physical models have explained
them all as single classes of events differing only quantitatively.

Similar questions have arisen with regard to Fast Radio Bursts (FRB).  If
they are all manifestations of the same processes, then any proposed
mechanism must be consistent with every FRB observed (allowing for
quantitative differences in parameters and viewing geometry), while if there
are two or more qualitatively different mechanisms or source objects
producing FRB this constraint is relaxed.  An apparent excess of bright
bursts over the power law distribution of sources uniformly but randomly
distributed in space (the familiar $N \propto S^{-3/2}$ relation of
extragalactic radio astronomy, where $N$ is the cumulative number with flux
greater than $S$) has been used \citep{K17} to challenge the assumption of
a statistically uniform spatial distribution, but this has been disputed
\citep{ME18} on the grounds of possible discovery bias of the very bright
FRB 010724.

More recently, the fluxes and fluences of many additional FRB, including
homogeneous data sets from Parkes, UTMOST, ASKAP and CHIME/FRB, have been
measured and the rotation measures (RM) of a number of FRB have also been
obtained.  Some of these datasets include outliers, and again raise the
question of whether the outliers indicate the presence of two (or more)
qualitatively different events classified as FRB, or of two qualitatively
different environments of FRB (possibly, but not necessarily, related to
different models of the FRB themselves).

In the simplest possible case, which arises {in many} extant data sets,
there is a single outlier.  When can we infer that it represents a different
class of object or event?  When there is only one outlier, cluster analysis
is {inapplicable}.  

The distribution of brightnesses of astronomical objects is a convolution of
their intrinsic luminosities or energies with their spatial distribution,
which is usually distributed over a very large range of distances.  {As
a result,} objects of distinct luminosity do not form distinct clusters of
apparent brightness.  For example, there are bright nearby dwarf stars and
dim distant giant stars; {even if stars were clustered in luminosity
they would not be clustered in apparent brightness.}

Each interval of intrinsic {luminosity} has a $N \propto S^{-3/2}$
distribution of observed flux or fluence $S$, as must the integral over
their luminosity distribution.  This only breaks down when a characteristic
scale enters, such as the characteristic distance to the nearest member of
the population (the $-1/3$ power of its {number} density), {recently
used by \citet{LBK} to estimate the number density of sub-classes of FRB,}
or when the statistically homogeneous Euclidean distribution is cut off by
the size of the Galaxy, the Local Group, other cosmic structure, or the
Hubble distance.   {A related method \citep{WVP} has been used to infer the
number of objects in a population from the total flux, the flux of the
brightest member, and a power law fit to their luminosity function.}

How extreme must an outlier be, in order that we may conclude, {with
statistical confidence,} that it is qualitatively different from objects or
events it otherwise resembles?  This question arises even when only one
parameter is studied and when as few as two objects have been observed.

{Sec.~\ref{PL} discusses the significance of power law distributions,
not only in astronomy, but more generally.  Sec.~\ref{SO} asks if a
single outlier in some variable is consistent with a power law distribution
of that variable.  This is motivated by the outlying flux and fluence of the
discovery FRB 010724, the rotation measure of FRB 121102, the flux and
fluence of FRB 200428, and the giant outbursts of soft gamma repeaters
(SGR).  These objects or events are single outliers, not clusters of data;
their consistency with a power law may be assessed by comparing them to the
second-most outlying object or event.  Sec.~\ref{A} applies this method to a
number of classes of astronomical objects and events, while the Discussion
(Sec.~\ref{D}) summarizes these results and their implications.}
\section{Power Laws}
\label{PL}
It is a trivial observation, known at least since Kolmogorov's work on
turbulence in the 1940's \citep{K41,T72,F95,B00}, and discussed by
\citet{K86} in the context of the model now called Self-Organized
Criticality, that a power law distribution of a parameter as a function of
an independent variable is inconsistent with the existence of a
characteristic scale of that variable: A power law is a straight line on a
log-log plot, any portion of which may be transposed to any other portion by
scaling factors.  Any deviation from a {single} power law defines a
characteristic scale, the value of the independent variable at which the
actual distribution deviates, {perhaps by a change of} the exponent in
the power law.

{Deviations from power laws may have many different origins.  These may
include selection biases.  The most obvious example of such a bias results
from the fact that more distant sources are fainter.  If the sources are
standard candles, there will be a cutoff distance beyond which none are
detected.  In this case, the deviation from the power law reflects their
detectability, not any intrinsic property of the sources.  In order to
ensure that a deviation from a power law indicates an intrinsic property,
such biases must be avoided.}

If there is no characteristic scale, so that all objects in the class are
qualitatively similar although quantitatively different, the observed
distribution must be a power law.  Hence consistency of the distribution
with a power law is a test of the hypothesis that the objects cannot be
divided into qualitatively different subclasses.  Inconsistency with a
power law demonstrates the existence of a qualitative distinction.  An
outlier, by definition, is inconsistent with the distribution observed of
the other objects.  The probability of this occurring by chance may be
calculated as a function of the power law slope of the other elements
of the class and of the degree to which the outlier is beyond the
extrapolation of the observed distribution of the other objects.  This is
most simply quantified by comparing the value of the observed parameter of
the most extreme (outlying) element to that of the second-most extreme.  If
there are multiple outliers it will generally be evident that there are two
distinct subclasses of objects, that can be separated by cluster analysis,
but in some cases of interest there is only one outlier.
\section{Single Outliers and First/Second Ratio}
\label{SO}
All power law distributions must have at least one cutoff or break in order
that the total number of objects be finite.  When the independent variable
is flux, fluence, luminosity, or an analogous energetic parameter, the break
must be sharp (between two power laws, their exponents must differ by $> 1$)
in order that both the number of objects and their integrated radiated power
or energy be finite.  The question is not whether the power law is broken
but whether the break occurs within the observed range of the independent
variable.  If it does, then the break divides the observed objects into two
qualitatively different classes.  If not, then all the observed objects {\it
may\/} be qualitatively similar, though quantitatively differing.  Even so,
they are not {\it necessarily\/} qualitatively similar, as shown by the
examples of dwarf and giant stars, or main sequence and white dwarf stars,
all of whose flux distributions are described by a $N \propto S^{-3/2}$ law
because they are homogeneously distributed in a Euclidean space (up to
cutoffs at the thickness of the Galactic disc and the distance of nearest
neighbors).

If there are many outliers, by definition of ``outlier'' a break divides
them from the remainder of the objects.  It will generally be evident that
they form a distinct distribution, often with other evidence of their
difference (for example, supernov\ae\ and nov\ae\ have very different
spectra and temporal behavior, while long and short GRB differ in duration).  

The problem addressed here is that of a single outlier.  When can we say
that there must be a break in the distribution between it and the object
with the second-most extreme value of the independent variable, that
observation of the outlier is inconsistent with a power law that describes
the distribution of the independent variable over less extreme objects?

Suppose a power law differential number distribution as a function of an
independent variable $x$
\begin{equation}
	{dN \over dx} = C x^{-\gamma}.
\end{equation}
By definition, there is one object with $x > x_2$, where $x_2$ is the
second-highest value of $x$.  Then
\begin{equation}
	1 = \int_{x_2}^\infty\!{dN \over dx}\,dx = {C \over \gamma -1}
	x_2^{-\gamma+1}
\end{equation}
and
\begin{equation}
	C = (\gamma-1)x_2^{1-\gamma}.
\end{equation}

The requirement that the total number of objects be finite, integrating as
$x \to \infty$ implies $\gamma > 1$.  If $x$ represents an energy-like
quantity (flux, fluence, luminosity, {\it etc.\/}), finiteness of the
integral implies $\gamma > 2$.  In homogeneously filled three dimensional
Euclidean space in the limit $x \to \infty$ $\gamma = 5/2$.  The divergence
\begin{equation}
	\int\!S\,dN = \int\!S{dN \over dS}\,dS \propto \int\!S^{-3/2}\,dS
	\propto S^{-1/2}
\end{equation}
as $S \to 0$ describes Olber's Paradox, resolved by a cutoff at large
distance (Hubble radius, or size of the Galaxy) and small $S$.  In two
dimensional space, such as the Galactic disc, $\gamma = 2$.  In practice,
$\gamma$ is found by fitting to the observed distribution for $x < x_2$.

The probability that the highest value of $x$ is as large or larger than
the extreme outlier $x_1$ is
\begin{equation}
	P(x \ge x_1) = \int_{x_1}^\infty\!{dN \over dx}\,dx = 
	\left({x_1 \over x_2}\right)^{1-\gamma}.
\end{equation}
An observed $x_1$ is inconsistent with an extrapolation of the power law at
a confidence level $1-P$ if
\begin{equation}
	\label{P}
	{x_1 \over x_2} > P^{1/(1-\gamma)}.
\end{equation}
If $\gamma$ is known, inverting this expression shows the minimum value of
$x_1/x_2$ required to reject, at a confidence level $1-P$, the hypothesis
that the power law is unbroken between $x_2$ and $x_1$.  Rejection implies
that the outlier differs qualitatively from the remainder of the objects.
If, as is often the case, $\gamma$ is not accurately determined, an observed
value of $x_1/x_2$ sets a coupled constraint on the value of $\gamma$ and
the confidence with which an unbroken power law can be rejected.

\begin{figure}
	\centering
	\includegraphics[width=0.99\columnwidth]{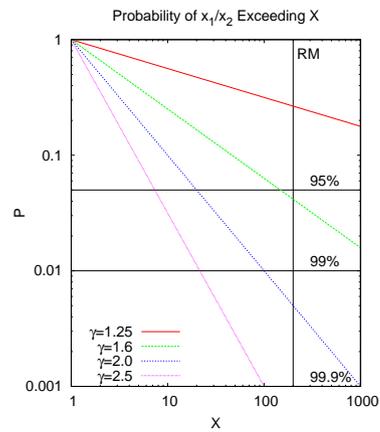}
	\caption{\label{Pplot1}The probability that $x_1/x_2$ exceeds $X$ for
	several values of $\gamma$.  Levels of statistical significance are
	indicated.  $x_1/x_2=200$ is the first/second ratio of FRB Rotation
	Measure.} 
\end{figure}

\begin{figure}
	\centering
	\includegraphics[width=0.99\columnwidth]{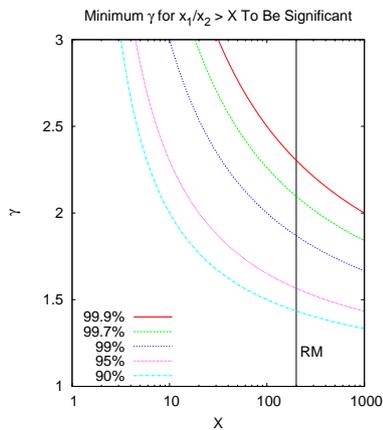}
	\caption{\label{Pplot2}The the minimum value of $\gamma$ as a
	function of $X$ for which $x_1/x_2 > X$ indicates that the outlier
	is inconsistent with an extrapolation of the power law, for the
	indicated levels of statistical confidence.  $x_1/x_2 = 200$ is the
	first/second ration of FRB Rotation Measure.}
\end{figure}

Fig.~\ref{Pplot1} shows $P$ as a function of $x_1/x_2$ for several values of
$\gamma$.  Fig.~\ref{Pplot2} shows $\gamma$ as a function of $x_1/x_2$ for
several values of statistical significance $1-P$.
\section{Applications}
\label{A}
The Table shows most extreme/second most extreme ratios for several
astronomical datasets.  {Most of these objects are variable, so the
values in the indicated catalogues are shown.  This introduces uncertainty
because there may be biases resulting from differences in observing
methodology and the selection of catalogued values; for example, stronger
variable sources may be observed more frequently and their greatest
strengths recorded in the catalogue, biasing their catalogued strengths
upward.  Extreme examples of this are the maximum observed fluxes reported
for transients.  Some catalogues, such as those for specific FRB surveys,
are likely homogeneous with minimal bias.

Objects distributed homogeneously
in three dimensional space have $\gamma = 5/2$, while Galactic disc objects,
such as most entries in X-ray catalogues, are expected to have $\gamma = 2$
because of the two dimensional geometry of a disc.  These values of $\gamma$
assume no cutoff resulting from the cosmological redshift and the finite
extent of the Galactic disc; these assumptions are tested here, and shown to
be likely invalid for the AGN and 3CR (extragalactic) catalogues, and (with
somewhat less significance) for the 4U catalogue.

Direct comparisons of the flux and fluence of FRB 200428 to those of other
FRB are not quantitative because FRB 200428 does not come from a homogeneous
catalogue.  It is included here to show how its extreme outlying position is
explained by its location in the Galaxy, a mass concentration inconsistent
with a homogeneous Universe.  The cosmological $\gamma = 5/2$ cannot be
extrapolated down to Galactic distances; $\gamma = 5/2$ is excluded with
very high confidence.  This is not a new discovery(!), but illustrates the
method.  It also lends confidence to the assertion that the Parkes catalogue
of FRB does not provide compelling evidence that the discovery FRB 010724 is
an outlier inconsistent with a statistically homogeneous cosmological
distribution.

The 3CR catalogue \citep{3CR,3CRprep} is a classic catalogue of radio
sources observed at 178 MHz.  It is used here because its sources have been
identified as Galactic \citep{S85} or extragalactic.   Many entries in the
4U X-ray source catalogue \citep{4U} have not been identified, but are
assigned as Galactic if $b \le 20^\circ$.  Fluences of SGR 1806$-$20
outbursts are from \citet{H05,P05,G07} with $\gamma$ from \citet{G00,G06}
($x_1$ and $x_2$ are not from a homogeneous catalogue, introducing
additional uncertainty).  Crab pulsar Giant Pulses are from \citet{BC19}.
FRB are from the homogeneous catalogues in \citet{H21} except for CHIME from
\citet{CFC} and FRB 200428 from \citet{B20,C20}.  FRB rotation measures (RM)
are from \citet{frbcat} and \citet{Hi21}; the value of $\gamma$ is for the
supernova remnant (SNR) model described in Sec.~\ref{D}.  FRB 121102 bursts
are from a single homogeneous five hour sample {with $x_1/x_2 = 1.55$,
the average of the values of \citet{G18} (1.75) and \citet{Z18} (1.35), but
with $\gamma = 1.7$ from \citet{L17,WZ19}}.}
\begin{table}
	\begin{center}
	\begin{tabular}{|crccc|}
		\hline
		Parameter&N&$x_1/x_2$&$\gamma$&Significance\\
		\hline
		Stars (V-band)&&1.94&5/2&63\%\\
		AGN (V-band)&&10&5/2&97\%\\
		3CR (extragalactic)&298&8&5/2&96\%\\
		3CR (Galactic)&38&8&2&87\%\\
		4U  (Galactic)&181&18&2&94\%\\
		4U  (transients)&12&3.5&2&72\%\\
		SGR 1806$-$20 fluence&760&$7 \times 10^4$&1.7&99.96\%\\
		Crab Giant Pulses&$>1100$&1.25&2.8&33\%\\
		FRB Fluxes (Parkes)&31&4.3&5/2&89\%\\
		FRB Fluxes (UTMOST)&15&1.37&5/2&38\%\\
		FRB Fluxes (ASKAP)&42&1.15&5/2&19\%\\
		FRB Fluxes (CHIME)&536&1.33&2.4&33\%\\
		FRB Fluences (Parkes)&31&1.1&5/2&17\%\\
		FRB Fluences (UTMOST)&15&1.71&5/2&55\%\\
		FRB Fluences (ASKAP)&42&2.1&5/2&67\%\\
		FRB Fluences (CHIME)&536&1.0&2.4&0\%\\
		FRB 200428 Flux&&$1.7\times 10^4$&5/2&>99.9999\%\\
		FRB 200428 Fluence&&$3.6\times 10^3$&5/2&>99.999\%\\
		FRB RM&19&200&5/4&73\%\\
		FRB 121102 Fluxes&93&1.55&1.7&26\%\\
		\hline
	\end{tabular}
	\caption{\label{Data}Ratios of most extreme to second most-extreme
	members of various astronomical datasets.  Where meaningful, $N$ is
	the number of data.  {Fractional and integral $\gamma$ are
	theoretical values; of the FRB catalogues, only CHIME contains
	enough data for a meaningful empirical $\gamma$, which is shown and
	is close to the theoretical value of 5/2.}  The final column is the
	significance of any inconsistency of $x_1/x_2$ with a single power
	law with an exponent known either theoretically or empirically from
	the distribution of less extreme members; lower values imply
	consistency.}
\end{center}
\end{table}
\section{Discussion}
\label{D}
Table 1 shows a number of significant outliers, of which the most extreme is
FRB 200428.  This is explained by its presence in the Galaxy whose density
(of mass, and likely of whatever astronomical objects make FRB) is orders of
magnitude greater than the mean density of the local Universe.  The
extragalactic $\gamma = 5/2$ does not allow for the Galactic density
concentration.  Because we are also located in the Galaxy, it is
unsurprising that the one Galactic FRB should be an outlier in flux and
fluence.  This is not news, but confirms that the method finds the obvious.

More interesting are the facts that the AGN 3C273 is a significant outlier
in the visible-light AGN population, Cyg A is a significant outlier in
the 3CR extragalactic (radio galaxy) distribution and Sco X-1 is a
significant outlier in the 4U Galactic 2--6 keV X-ray source distribution.
Perhaps these objects are fundamentally different from other members of
their classes.  The extragalactic FRB distributions contain no such
outliers, attesting to the homogeneity of their populations; even the
discovery FRB 010724 is not a significant outlier compared to {the
second-brightest Parkes} FRB 110214.
It is no surprise that this analysis confirms that the giant December 27,
2004 outburst of SGR 1806-20 \citep{H05,P05} is a significant outlier,
inconsistent with extrapolation of its lesser flares.  Such giant outbursts
are naturally explained as a global reordering of a magnetic field much
greater than those of radio pulsars, as suggested for the March 5, 1979
outburst of SGR 0525-66 \citep{K82} while less energetic outbursts may be
analogous to Solar flares; this is now known as the ``magnetar'' model.

The distribution of FRB RM, with $x_1/x_2 \approx 200$, presents a different
problem because there is no {\it a priori\/} expectation of {the
exponent} $\gamma$.  Eq.~\ref{P} indicates that $\gamma > 1.43$ can be
excluded with 90\% confidence, $\gamma > 1.57$ with 95\% confidence and
$\gamma > 1.87$ with 99\% confidence.  In one simple model the rotation
measure is predicted to arise from an expanding SNR.  In such an object the
magnetic field would be expected to scale $B \propto R^{-2} \propto t^{-2}$,
the electron density to scale $n_e \propto R^{-3} \propto t^{-3}$ and
$\text{RM} \propto B n_e R \propto R^{-4} \propto t^{-4}$, where $t$ is the
age.  The irregular fluctuations of the observed RM of FRB 121102
\citep{Hi21} are inconsistent with such a simple model, but it might
approximately describe the scaling.  If the likelihood of detection does not
vary rapidly with time, then the cumulative number $N$ of FRB with ages $\le
t$ is proportional to $t$ and $\gamma = 1.25$.  Surprisingly, there is not
strong statistical evidence that FRB 121102 is an outlier from this model.
It would be an outlier from models that indicate a steeper dependence of $N$
on RM.

{Table~\ref{Data} shows that the homogeneous FRB catalogues are
consistent with single power law distributions.  The AGN and 3CR
(extragalactic) catalogues are not, suggesting some natural scale of
activity or bias.  A possible source of bias in observations of these
variable objects is more frequent observation of the brightest, so they
are more likely to be observed in more extreme and luminous states.  This
is a cautionary tale.

The giant outburst of SGR 1806-20 is definitely inconsistent with an
extrapolation of the brightness distribution of its lesser bursts; it is
qualitatively different.  This is not a new result, but now is quantified.
The extraordinary nature of FRB 200428 is hardly a surprise; Galactic
objects are much closer than extragalactic objects, and would be expected
to be much brighter, an expectation confirmed with high confidence.  There
is also a less trivial conclusion: the Galaxy does not contain large numbers
of micro-FRB, weaker sources of coherent radio radiation than FRB 200428,
but still readily observable because of their proximity; FRB 200428 was
orders of magnitude above the detection thresholds of the instruments
\citep{B20,C20} that discovered it.  {See also the discussions of
\citet{L20,M20}.}  Finally, in at least one model, the extraordinary RM of
FRB 121102 is not so extraordinary at all.}
\section{Data Availability}
This theoretical study generated no new data.

\label{lastpage} 
\end{document}